\documentclass[11pt,fleqn]{article}
\usepackage{amsmath}
\usepackage{amssymb}
\usepackage{times}
\usepackage{mathptm}
\usepackage{ifthen}
\usepackage{graphics}
\usepackage{color}
\usepackage{subfigure}
\usepackage{epsfig}
\usepackage[pdftex,colorlinks,linkcolor={blue},citecolor={blue}]{hyperref}
\usepackage[margin=1in]{geometry}
\usepackage[T1]{fontenc}
\usepackage[authoryear,round]{natbib}
\usepackage{titlesec}
\usepackage{appendix}
\usepackage{enumitem}
\usepackage{color}

\long\def\comment#1{}

\begin{document}

\title{Making intersections safer with I2V communication}
\author{Offer Grembek, Alex Kurzhanskiy, Aditya Medury, Pravin Varaiya and Mengqiao Yu\\
University of California, Berkeley\\
\{grembek, akurzhan, amedury, mengqiao.yu,varaiya@berkeley.edu\} }
\maketitle
\begin{abstract}
Intersections are hazardous places. Threats arise from interactions among pedestrians, bicycles and vehicles,  more complicated  vehicle trajectories in the  absence of lane markings,  phases that prevent determining who has the right of way, invisible vehicle approaches, vehicle obstructions, and illegal movements. These challenges are not fully addressed by the ``road diet'' and road redesign  prescribed in Vision Zero plans, nor  will they be completely overcome by autonomous vehicles with their many sensors and tireless attention to  surroundings.
Accidents can also occur because drivers, cyclists and pedestrians do not have the  information they need to avoid wrong decisions.  In these cases, the missing information can be computed and broadcast by an intelligent intersection. The information gives the current full signal phase, an estimate of the time when the phase will change, and the occupancy of the blind spots of the driver or autonomous vehicle. The paper develops a design of the intelligent intersection, motivated by the analysis of an accident at an intersection in Tempe, AZ, between an automated Uber Volvo and a manual Honda CRV and culminates in a proposal for an intelligent intersection infrastructure. The intelligent intersection also serves as a software-enabled version of the `protected intersection' design to improve the passage of cyclists and pedestrians through an intersection.
\end{abstract}
\section{Introduction}\label{sec1}
4,000 New Yorkers are seriously injured and 250 are killed each year in traffic crashes. This is the leading cause of injury-related death for children under 14, and the second leading cause for seniors. In 2014 Mayor Bill de Blasio launched New York's Vision Zero Action Plan seeking to eliminate these injuries and deaths \citep{NYC_VZ}. 
In the San Francisco Bay Area  fatalities from crashes increased  43 percent between 2010 and 2016 to reach 455 killed, of which 62 percent were bicyclists or pedestrians \citep{MTC_VZ}. 
In California, Vision Zero (VZ) plans have been initiated in Los Angeles, Sacramento, San Mateo, San Jose, Santa Barbara, San Francisco and San Diego.

Autonomous vehicles (AVs) offer a radically different path to a crash-free urban road network. The widespread deployment of AVs, with their many sensors and tireless attention to their surroundings, is expected to eliminate all 94 percent of crashes involving  human  error, with no sacrifice of mobility \citep{waymo_safety, GM_safety2}.  

VZ plans focus on dangerous intersections. VZ investments physically modify intersections to limit vehicle mobility through lane elimination and enforced speed reduction, sidewalk extensions that shorten pedestrian crossings, and dedicated bike lanes.  The aim is to eliminate serious accidents and deaths by restricting vehicle movement while facilitating walking and biking.

Intersections also pose a challenging environment to AVs because of the complex interactions among pedestrians, bicycles and vehicles, more complicated vehicle trajectories in the absence of lane markings to guide vehicles, split signal phases that prevent determining who has the right of way, invisible vehicle approaches, and illegal movements.  

The concentration of VZ investments on  intersections is to be expected since 50\% of all injury/fatal crashes occur at or near an intersection \citep{FHWA_intersection}. The initial safety outcomes of VZ investments were uncertain.  An evaluation of 12 NYC streets that received a `slow zone' treatment during 2012-14 found ``Overall, the number of wrecks remained virtually unchanged on those problem corridors, and casualties fell only 4\%.'' Six of the 12 streets reported a reduction in accidents with injuries and fatalities and six reported an increase \citep{NYVZ}. However, the most recent report shows that rigorous enforcement against speeding, texting while driving and failure to yield to pedestrians,  combined with street design improvements including Leading  Pedestrian Intervals (LPIs), addressing left turns at intersections and building out the protected bike lane network, led to a 45\% decline in pedestrian fatalities from 2013 to 2017 \citep{NYC_VZ}.

In Fall 2016 the City of Berkeley redesigned the  intersection at Hopkins Street and The Alameda. This `protected intersection' now includes four concrete `refuge islands' that narrow the formerly wide intersection, so cars must slow down to negotiate a smaller turning radius. The islands  guide bicycles as they approach the intersection and back towards the traffic lane afterward. Pedestrians have a shorter crossing distance. The result is mixed. Drivers have complained about the increased difficulty turning right. Tire marks have obscured the paint on the outside of some of the islands, as drivers turn too sharply. At one corner tire marks record drivers unwilling to wait in a queue to turn right, and go into the bike lane instead. A citizen who has been crossing this intersection for 17 years said that ``squeezing the traffic'' into a narrower space makes this ``a stressful intersection'' and irritates 
drivers \citep{BerkeleyVZ}.

VZ plans in Edmonton, Canada have had  success. Over a six-year period strict enforcement against speeding and red light violations resulted in a 12\% drop in collisions, and implementing a 30km/h speed zone at all elementary schools led to a 4\% reduction in serious collisions and speed reduction of 12 km/h.  Engineering improvements include left-turn-only green flashing arrow on traffic lights and pedestrian crossing upgrades.  The city has also raised community awareness in traffic safety \citep{EdmontonVZ}.  

Autonomous vehicles (AV) offer a radically different path to a crash-free city. The widespread deployment of AVs, with their precomputed street maps, on-board sensors, and tireless attention to the sensor readings, is expected to eliminate all 94 percent of crashes a to human error, with no sacrifice of mobility.\footnote{The 94\% is misleading.  The NHTSA report, based on 2005-2007 data, states the assignment is not intended to blame the driver for causing 
the crash \citep{NMVCCS}.}  The claim has captured public attention and secured unlimited venture capital. But the claim is unverifiable at present. In 2016, Waymo reported that its safety drivers disengaged autonomous driving once every 5,000 miles; in 2017 this rate was  5,500 miles. Waymo reports a disengagement only when it identifies the event as having ``safety significance and should receive prompt and thorough attention from our engineers in resolving them'' \citep{CADMV}. Safety significance seems to mean that without the disengagement, there could have been a crash. Waymo's reported 5,500 miles per disengagement may be compared  with the 500,000 miles per reported accident in the U.S. today.

Advanced AV designs rely exclusively on their sensors and maps and do not communicate with the infras- tructure or with other vehicles. However, one auto parts supplier has expressed interest in systems with infrastructure sensing and communication  and one manufacturer has announced plans for a connected car platform \citep{Continental, fordcc}. In any event, intersections pose a challenging environment to AVs because of the complex interactions among pedestrians, bicycles and vehicles, more complicated vehicle trajectories in the absence of lane markings to guide vehicles, split signal phases that prevent knowing who has the right of way, invisible vehicle approaches, and illegal movements. AV test reports do not disclose the number or difficulty of intersections traversed in autonomous mode.

 The focus on intersection safety in AV design is also to be expected since, between October 2014 and April 2018, 58 out of 66 or 88 percent of self-reported AV crashes in California have occurred near intersections \citep{CADMV_Acc}.   Moreover, a study of these reported accidents also casts doubts on AV capability to drive collision-free in intersections, at least in the near future \citep{favaro}.   On the other hand, proposals for collision-free intersections, e.g. \citep{safety-ahn,Malikopoulos} presuppose a highly centralized communication and control and 100 percent AV market penetration. which will not be realized in the foreseeable future, except in highly restricted environments.

Reported VZ plans and  AV tests make no use of the opportunities made possible by infrastructure sensing and I2V communication that make movement in intersections safer for automobile drivers, cyclists and pedestrians by reducing the spatial and temporal uncertainty they face.  VZ plans appear to be limited to modification of intersections that constrict vehicle movement, with no role for these opportunities.   AV systems all seek to achieve self-driving with no assistance from the infrastructure \citep{waymo_safety, GM_safety2}.

This paper argues the case for exploiting the opportunities offered by an intelligent infrastructure.  As the penetration of connected and automated vehicles (CAV) increases, I2V can improve safety even further.  Section \ref{sec2} explains how intersections are unsafe; section \ref{sec3} analyzes a revealing accident that could have been avoided by I2V communication; section \ref{sec4} describes an approach to using I2V communication to make intersections safer and section \ref{sec4a}  proposes the intelligent intersection based on this approach.  Section \ref{sec5} shows how this approach would avoid the accident of section \ref{sec3}. Section \ref{sec5a} illustrates steps in a procedure that automates key steps in this approach. 
Conclusions are collected in section \ref{sec6}.

\section{Why intersections are unsafe} \label{sec2}
Unlike streets with well-defined lane dividers, intersections do not have markers in the pavement that separate users and movements.  The paths of vehicles, bicycles and pedestrians cross each other within intersections, creating `conflict zones' and the potential for crashes.  So avoidance of crashes requires the movements of different agents to be separated in time and space.  It is impossible to fully achieve this separation and so the risk of intersection crashes remains.

Traffic signal control provides limited separation because it does not simultaneously give the right-of-way (green light) to two conflicting movements.  Although critical to safety, its effectiveness is often compromised.  A driver (or autonomous vehicle) planning a certain movement (say a left turn) can see from the signal light whether her own planned left turn is permitted, but she may be unable to figure out whether another conflicting movement (say a through movement by another vehicle or bicycle, or a pedestrian crossing) is also permitted.  That is, the signal light visible to the driver does not provide the complete phase information.  Similarly, a pedestrian or cyclist undertaking a  movement may be unaware of a permitted conflicting vehicle movement.  This \textit{spatial uncertainty} about rights-of-way  can be eliminated if the infrastructure provides the complete phase information.  

Furthermore, road users do not rely solely on their current view of the  traffic signal.  They also predict how the signal will change in the next few seconds.  An accurate prediction of the duration of the current phase and the upcoming phase can be supplied by the intersection, thus reducing \textit{temporal uncertainty} about rights-of-way.  This information
can be provided by processing signal phase data accumulated at the intersection.  The information is part of the SPaT (signal phase and timing) I2V message that the SAE has standardized \citep{J2735}.  Intersections could broadcast SPaT messages \citep{SPATDileep}.

Even when the driver (cyclist or pedestrian or AV) knows the complete phase, her knowledge of the intersection state will be limited by the extent to which her view of the intersection is obscured by other users.  If the driver cannot fully see a conflict zone, she must guess  whether there is a hidden user undertaking a movement in the conflict zone.  This dilemma can lead either to slow driving that is overly cautious (when there is no hidden user), or to driving optimistically at a normal speed with greater risk of a crash (because there is a hidden user).  The intersection infrastructure can process sensor measurements to determine the presence or absence of a hidden user and communicate that to the driver, thereby eliminating the risk of an overly pessimistic or optimistic assessment \citep{ITSWC2017}.

Lastly, even when no conflicting movement is present, a crash may occur from the illegal movement of  another car, bicycle or pedestrian.  A common example is a car or bicycle running a red light or a pedestrian crossing against a `dont walk' signal.  If appropriate sensor measurements can be acquired and processed rapidly, the driver could be warned to take evasive action.

In summary, sensor data acquisition and processing at an intersection can provide I2V messages that give complete phase information, predict the signal phase and timing in the next cycle, accurately assess potential conflicts, and warn of the danger from traffic violators.  Call an intersection with this capability an intelligent intersection.  Upgrading to an intelligent intersection is not cheap.  However, data collected at an intersection can be analyzed to estimate how many crashes can occur.  Intersections can be ranked accordingly and  investment can be directed at the most unsafe intersections.

\section{An accident} \label{sec3}
\begin{figure}[h!]
\centering
\includegraphics[width=6.5in]{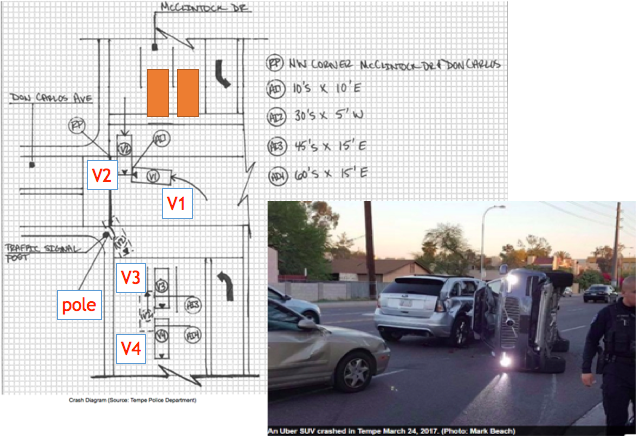}
\caption{Left: diagram from the police report of an intersection where a Honda CRV (V1) traveling north made a left turn and collided with an Uber automated Volvo (V2) traveling south at 38 mph in a 40 mph zone. After the collision, the Uber Volvo hit a signal pole, and two other vehicles (V3, V4),
shown in the inset.  Right: the accident caused heavy damage but no one was seriously injured.  Source: \citep{UberAcc}.}
\label{fig-fig1}
\end{figure}

The accident described in this section occurred on March 24, 2017.  See Figure \ref{fig-fig1}.
Vehicle V1 (Honda CRV) northbound in the left turn lane of S. McClintock Dr entered the intersection during green, with 5 s left in the crosswalk timer, stopped, made a slow left turn onto E. Don Carlos Ave, and collided with vehicle V2 (Uber automated Volvo), southbound in lane 3 of S. McClintock Dr, which entered the intersection on yellow at 38 mph (56 fps).

After being hit, the Volvo continued across the intersection, struck a traffic signal pole,  flipped on its side and collided with Vehicles V3 (Hyundai EST) and  V4 (Ford Edge), which were stopped in traffic southbound in lane 2 of S. McClintock Dr.  The self-driving Uber had the right of way and was programmed to enter the busy intersection at the speed limit while the light was  yellow, but a human driver likely would have slowed down. A witness told the reporter, ``All I want to say is it was good on the end of the [Honda] driving toward us, it was the other driver's fault [Uber] for trying to beat the light and hitting the gas so hard'' \citep{UberAcc2}.

Four possible errors contributed to the accident.  The Uber automated Volvo (V2)
\begin{enumerate}[noitemsep,topsep=0pt]
\setlength{\itemsep}{0in}
\item may not have known that traffic in the opposing direction was permitted to turn left;
\item did not predict that the light would turn yellow before it entered the intersection;
\item did not consider that the vehicles stopped  in the adjacent lanes 1 and 2 prevented it from seeing a left-turning vehicle until the Uber was within 10 feet of the stop bar and at a speed of 56 fps it could not come to a stop within 10 feet. (At a deceleration of 32 f/s$^2$, the Uber would stop in 49 ft.)
\end{enumerate}
The Honda (V1) driver
\begin{enumerate}[noitemsep,topsep=0pt]
\setcounter{enumi}{3}
\item was occluded by vehicles stopped  in lanes 1 and 2 and could only  see  10 feet into lane 3 and seeing no vehicle there, concluded that none was going to cross the intersection; she did not realize  that the occlusion from  stopped vehicles prevented her from seeing a vehicle more than 50 feet away approaching at 40 mph.
\end{enumerate} 
Errors 1 and 2 could easily be prevented by a SPaT message that gives the current phase and predicts when it will end \citep{SPATDileep}. Error 3 could be prevented by a calculation of the `blind spot' due to the occlusion from the vehicles stopped in the adjacent lane, together with an intersection collision avoidance (ICA) message.  Error  4 is difficult to avoid but it could be prevented by a warning sign (Signalized Left Turn Assist System) proposed in the CICAS program \citep{cicas,pathcicas} or by an intersection collision avoidance (ICA) message.  The ICA messages could be triggered by strategically placed sensors within the intersection as described in Section \ref{sec5}.

Notice that errors 1-4 above are due to insufficient information, which places the drivers in a dilemma: should they be optimistic and proceed at a normal speed and risk an accident, or
should they be pessimistic and slow down or stop.  

\begin{figure}[h!]
\centering
\includegraphics[width=6.5in]{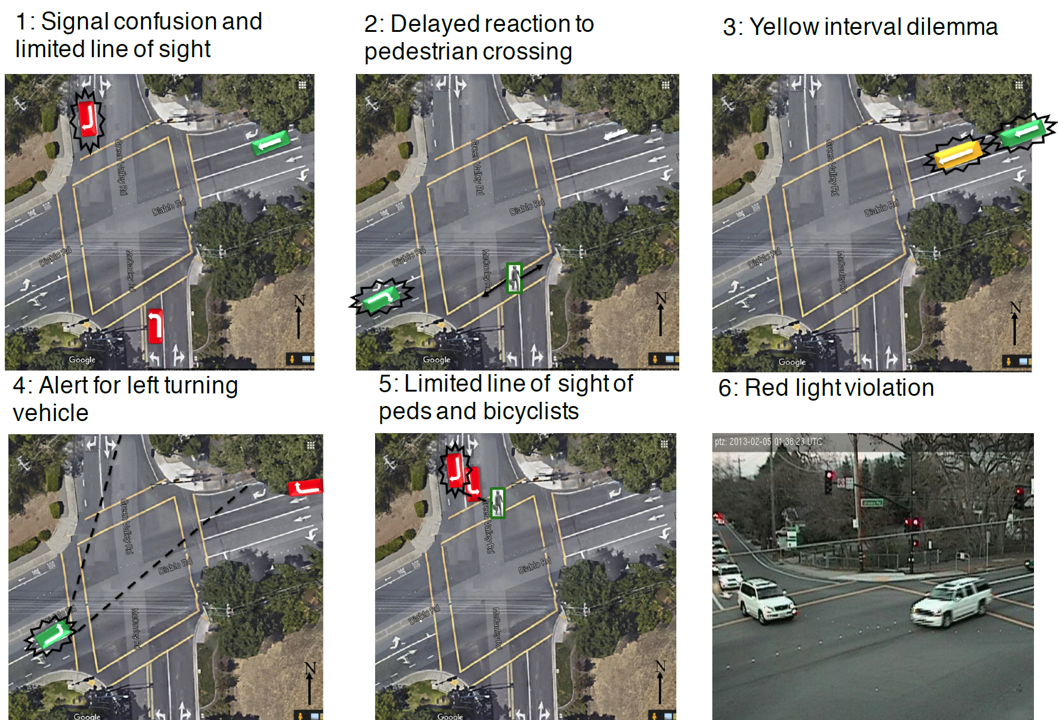}
\caption{Common intersection conflict scenarios.  Source: \citep{ITSWC2017}.}
\label{fig-fig2}
\end{figure}

The accident described above is one of several scenarios that pose dilemmas and induce errors of judgment on the part of intersection drivers \citep{ITSWC2017}.  Six common scenarios are described below:
\begin{enumerate}[noitemsep,topsep=0pt]
\setlength{\itemsep}{0in}
\item RTOR (right turn on red) signal phase confusion and limited LOS (line of sight):  RTOR vehicle can not determine if opposing traffic has the right of way;
\item delayed reaction to pedestrian crossing: RTOG (right turn on green) vehicle  needs a couple of seconds to detect pedestrian walking direction; 
\item yellow interval dilemma: the following through vehicle does not know when the traffic signal will turn yellow, which might trigger a rapid response from the leading vehicle;
\item left-turn alert: LTOG vehicle cannot detect the right turn vehicle; both share the same lane, creating conflict;
\item limited LOS for pedestrians/bicyclists: vehicle waiting to turn left blocks the LOS of RTOR vehicle, so it cannot see the  pedestrian on the crosswalk;
\item collision with red light violator.
\end{enumerate}
Figure \ref{fig-fig2} illustrates  the six scenarios, all in the same intersection; however, the red-light violation scenario is captured by a  camera at a different location.

\section{An  I2V system for intersection safety}\label{sec4}

The four-step approach  developed in this section for an I2V information system   makes intersections safer.  In step 1,   trajectories of users (cars, bicycles, pedestrians) are grouped into `guideways' corresponding to their  movements within an intersection.  In step 2,  `conflict zones' are identified as regions where two guideways intersect, creating the potential for an accident.  In step 3,   a procedure is used to determine if a planned movement can be safely executed with the information made available to the user.   This information consists in what  users  themselves can sense of the other users in the intersection, together with the SPaT message from the intersection.  The message gives the full current phase and an estimate of the time when the phase will change.
Most conflicts are resolved by step 3.
The conflicts that remain are due to blind zones.  In  step 4, sensor information provided by the intersection tells whether these blind zones are occupied by other users or not. Note that steps 1 and 2 are conducted offline, wheres steps 3 and 4 require real time information.
 Section \ref{sec5} illustrates this  four-step approach for the Uber accident described in section \ref{sec3}.

\begin{figure}[h!]
\centering
\includegraphics[width=6in]{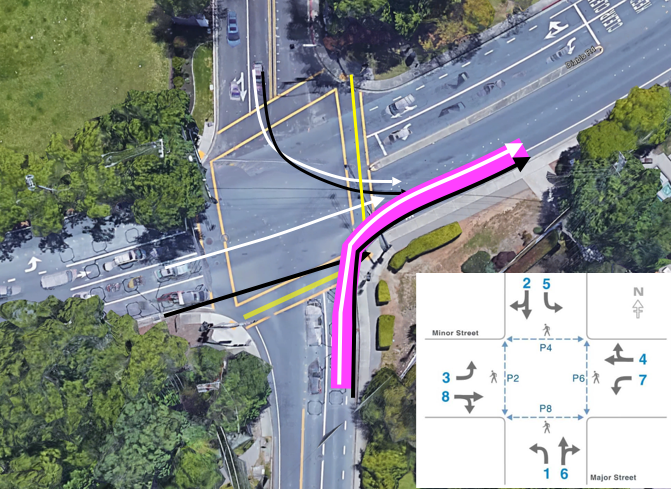}
\caption{A vehicle's right turn trajectory from the south in white.  The guideway of all right turn trajectories is in pink.  Seven trajectories can conflict with the right turn:  two other  vehicle trajectories in white,  three  bicycle trajectories in black, and  two  pedestrian trajectories in yellow.  Trajectories and guideways of non-conflicting movements are not shown.  The inset diagram defines the vehicle and bicycle phases $\phi1, ..., \phi8$ and the pedestrian phases P2, P4, P6, P8.}
\label{fig-fig3}
\end{figure}

The approach is described for a standard four-leg intersection.  Upon entering  any leg, a vehicle may turn left, turn right, or go straight, giving  in all 12  vehicle movements or phases. The eight non-right turn vehicle movements are numbered  1 through 8  and denoted   $\phi 1, \cdots, \phi 8$, as in the inset of Figure \ref{fig-fig3} \citep{stm}. The signal  lights control which phase is \textit{active} or actuated, i.e.\ which  movements have the green light. Right turn phases are not numbered, because it is assumed they are always permitted. 
Pedestrians can  only use crosswalks, so they have four   movements,  phases P2, P4, P6, P8, parallel to $\phi2, \phi4, \phi6 , \phi8$.  Pedestrian movements are regulated by the `walk/dont walk' signal, simultaneously with the corresponding phases, so P2 gets `walk' or `dont walk' at the same time that $\phi2$ gets `green' or `red', etc. Bicycles move alongside vehicles, so they have 12 movements as well, actuated concurrently with the corresponding vehicle movements.  (This is a simplification for ease of exposition: pedestrian, bicycle and vehicle movements need not be concurrent.)

Figure \ref{fig-fig3} is used to describe the approach.

\textbf{Step 1. Construct guideways.}  A trajectory is a path traced out by a vehicle as it moves through the intersection.  (Only  legal or permissible trajectories are considered.)  A \textit{guideway} is the bundle of vehicle trajectories that make the same movement.  A guideway starts  from  a single  lane entering the intersection and ends in a single outgoing lane.   There are  12 vehicle guideways corresponding to the 12 phases.  For the rest of this section we focus attention  on the  single white  trajectory of the vehicle making a right turn from the south in Figure \ref{fig-fig3}.  Call this the ego vehicle.   Its trajectory is inside the pink guideway of all right turn trajectories.  

The figure  shows two other vehicle trajectories in white, one making a left turn from the north ($\phi5$), the other making a through movement from the west ($\phi8$).  
Guideways for bicycles are adjacent to those for vehicles and  the figure displays three bicycle trajectories in black, one making  a right turn from the south, another making a through movement from the west ($\phi8$), and the third making a left turn from the north ($\phi5$). Pedestrian trajectories are confined to the crosswalks, which form the four pedestrian guideways.  Two pedestrian trajectories in yellow are shown (P6 and P8).  No bicycle or pedestrian guideway is shown.  We  will determine the information needed to make the ego vehicle's right turn movement safe.  The other movements are analyzed similarly.

Guideways may be  mathematically specified or empirically constructed.  Mathematically, a guideway for a particular movement  comprises all possible paths joining its entering and exiting lanes, constrained by a reasonable curvature.  The empirical construction of a guideway would use GPS traces or videos capturing vehicles or bicycles making a particular movement.  

\begin{figure}[h!]
\centering
\includegraphics[width=6in]{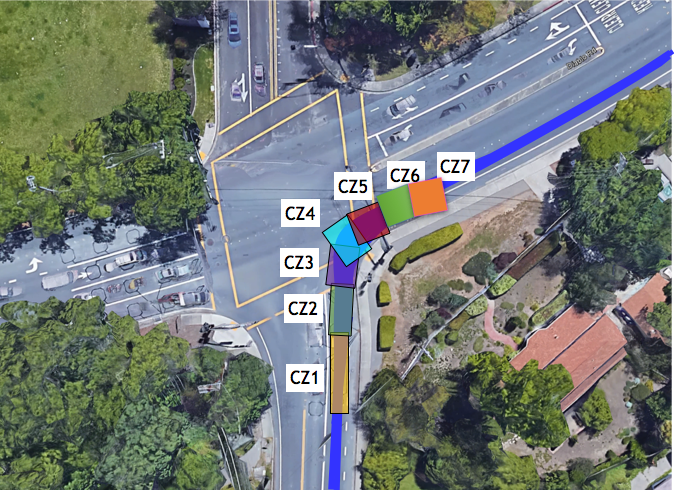}
\caption{The intersection of the seven guideways with the pink guideway yields seven conflict zones, CZ1, $\ldots$, CZ7. The conflict zones are shown disjoint for clarity, although in fact they overlap.}
\label{fig-fig4}
\end{figure}

\textbf{Step 2. Identify conflict zones.}  The seven trajectories -- two  vehicle,  three bicycle and two pedestrian -- all intersect the trajectory of the ego vehicle. The intersections of their guideways with the pink guideway identify the seven conflict zones that the ego vehicle must cross. From the  inset of Figure \ref{fig-fig3} one can see that the
following describe all  the  conflict zones involving the ego vehicle:

\noindent
\begin{enumerate}[noitemsep,topsep=0pt, label={CZ\arabic*}]
\setlength{\itemsep}{0in}
\item Conflict with right turn bicycle from south;
\item Conflict with pedestrian  on south crosswalk (P8);
\item Conflict with through vehicle from west ($\phi8$);
\item Conflict with through bicycle from west ($\phi8$);
\item Conflict with left turn bicycle from north ($ \phi5$);
\item Conflict with left turn vehicle from north ($ \phi5$);
\item Conflict with pedestrian on east crosswalk (P6).
\end{enumerate}
The conflict zones CZ1, $\cdots$, CZ7 can all be calculated ahead of time from a map of the intersection and the guideways.  They are shown in Figure \ref{fig-fig4} as disjoint rectangles for clarity, although in fact they overlap.  The ego vehicle must safely cross all seven conflict zones.

\textbf{Step 3.  Resolve conflicts.} 
There are three parts to this step which determines which of the seven conflicts can be  resolved.    Resolving a conflict means that the ego vehicle can determine whether or not another user will occupy a conflict zone at the same time as the ego vehicle, resulting in a crash.  If the ego vehicle determines that another user \textit{will}  occupy a conflict zone simultaneously with the ego vehicle, we assume the latter will use a collision avoidance procedure to avoid the accident; if the ego vehicle determines the conflict zone is unoccupied, it ignores this conflict zone.  Collision avoidance may simply require slowing down or speeding up without changing the ego vehicle's path  \citep{safety-ahn}.

\textit{Part 1. Using  signal light visible to ego vehicle.}   Some of the seven conflicts can be resolved by considering the signal light as seen by the ego vehicle, using the fact that two conflicting movements  never simultaneously have the green light.  Since the signal light may be red or green, the planned movement is either RTOR (right turn on red) or RTOG (right turn on green).

(i) Suppose this is a RTOR movement.  So phases $\phi6$ and $\phi1$ face a red light and P6 has `dont walk' signal.  Hence  conflict CZ7 cannot occur, but CZ1, CZ2, CZ3, CZ4, CZ5, CZ6  remain unresolved.

(ii) Suppose this is a RTOG movement.  So phase $\phi6$ faces a green light and P6 has `walk' signal,  phase $\phi8$ faces a red light and P8 has `dont walk' signal.  Hence conflicts CZ2, CZ3, CZ4 cannot occur, but CZ1, CZ5, CZ6, CZ7  remain unresolved.

\begin{figure}[h!]
\centering
\includegraphics[width=6in]{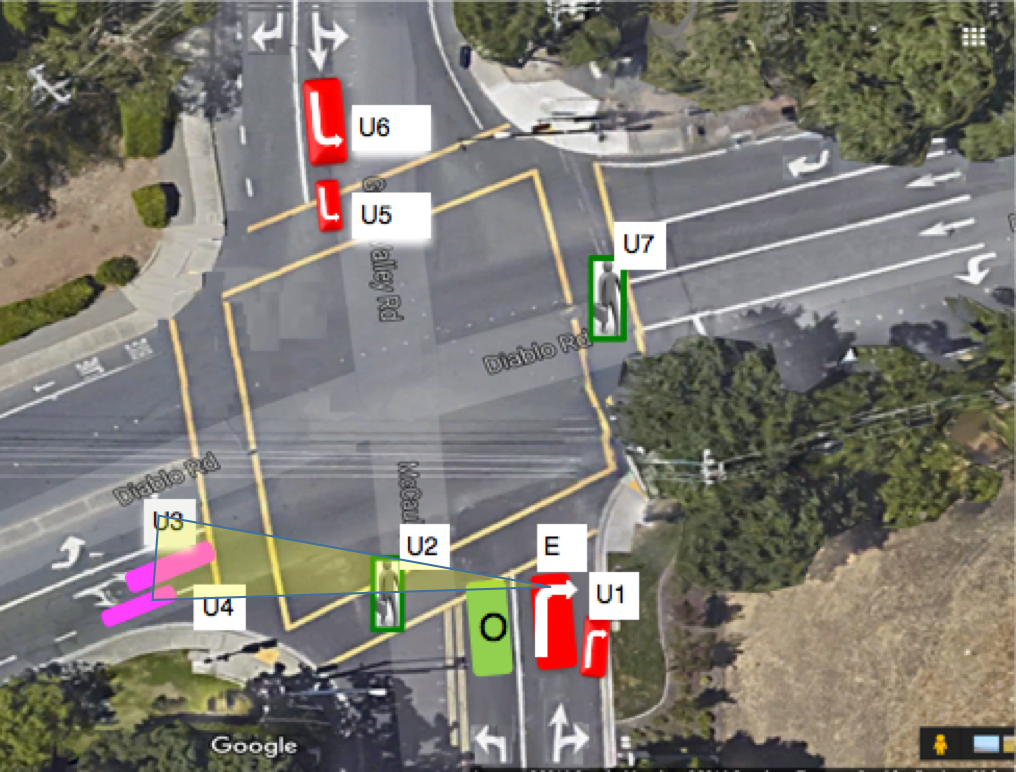}
\caption{The ego vehicle  must determine which of conflicts CZ2, CZ3, CZ4, CZ5 can be eliminated from what it sees of the intersection.   O obscures the triangular region from the ego vehicle's field of view so  it cannot  see U2, U3, U4, hence CZ2, CZ3, CZ4 remain.}
\label{fig-fig5}
\end{figure}

\textit{Part 2. Using ego vehicle's intersection view.}  The vehicle must decide from its view of the intersection which of the remaining conflicts can be resolved.   Figure \ref{fig-fig5} shows a configuration of other potential users in the intersection.  (The following analysis must be carried out for the prevailing configuration.)  E is the ego vehicle making a right turn.  U1, $\cdots$, U7 are the other users  whose movements    conflict with E:  U3, U6 are vehicles, U1, U4, U5 are bicycles, and U2, U7 are pedestrians. O is a vehicle stopped in the left turn lane next to E and obstructs E's view so E cannot see U2, U3, U4 (if they are indeed present) but can clearly see U1, U5, U6, U7.  
So E can resolve CZ1, CZ5, CZ6, CZ7.  However,   O prevents E from seeing whether or not U2, U3 and U4 are in fact present and pose a threat, so conflicts CZ2, CZ3, CZ4 remain.

(i) Suppose E is making a RTOR movement.  Then the unresolved conflicts are CZ2, CZ3, CZ4.

(ii) Suppose E is making a RTOG movement.  Then all the conflicts are resolved since U2, U3, U4 cannot move, and E can  safely complete the right turn.

\textit{Part 3.  Using signal phase and timing (SPaT) information.}  An equipped intersection  broadcasts a SPaT message every 100ms.  The message consists of the complete signal  phase (i.e.\ the signal phases faced by users at all legs) and an estimate of the time when the phase will change.  (In an actuated signal phase durations are not  deterministic, and  must be estimated \citep{SPATDileep}.)  We now see how SPaT information can help resolve conflicts.  

\begin{figure}[h!]
\centering
\includegraphics[width=6in]{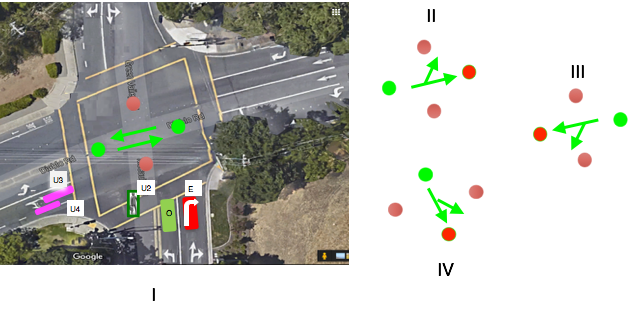}
\caption{The possible signal light configurations (I-IV) compatible with RTOR movement of the ego vehicle.}
\label{fig-fig6}
\end{figure}

Suppose E is making a RTOR movement and cannot resolve CZ2, CZ3, CZ4 because of obstruction by O.   The red signal seen by the ego vehicIe is compatible with the four possible signal light configurations shown in Figure \ref{fig-fig6}.  The ego vehicle does not know which configuration prevails, but discovers it from the SPaT message.
In configurations I and II West-East movement is permitted, so U2, U3 and U4 all can move and E cannot resolve CZ2, CZ3 and CZ4.  In configuration III, only U2 can move, so  E cannot resolve CZ2.  In configuration IV
U2, U3 and U4 cannot move, so all the conflicts are resolved, and E can complete the right turn.  

Thus upon receiving the SPaT message, it remains for E to resolve either conflict U2 (in configuration III) or conflicts U2, U3 and U4 (in configuration I).  This is done in Part 4.

The SPaT message will also  tell the ego vehicle that its signal will change to green in time $T$.  So after $T$, the vehicle's movement will automatically change from RTOR to RTOG and, as we have seen, all conflicts will be resolved.  $T$ may be as long as the cycle time, up to 2 mins.  The ego vehicle may decide it is worth waiting for time $T$  to complete its movement. 
Many AVs  today are programmed not to engage in RTOR movements, e.g. \citep{uberrtor,drive_ai}.
Waymo AVs reportedly have difficulty with unprotected left turns \citep{WaymoTurns}. 

\textbf{Step 4. Detect obstacles in blind zones.}  

This step is needed only if not all conflicts are  resolved and E decides to continue with the RTOG movement.
Suppose configuration I prevails so U2, U3 and U4 all can move and E cannot resolve CZ2, CZ3 and CZ4.
To continue with the right turn, E must be assured by infrastructure sensing that U2, U3  and U4 are not going to move to CZ2, CZ3 and CZ4 respectively before E moves.  Observe that U2, U3 and U4 are all upstream of their
corresponding conflict zones along their guideways.  Suppose E will take time $\tau$ to traverse the furthest conflict zone (CZ4 in
Figure \ref{fig-fig4}).  Call blind zone BZ$i$ that part of the guideway from which U$i$ can reach 
CZ$i$ in time $\tau$.  (This is the `backward reachability set' of U$i$ in time $\tau$.)  The obstacle O has created these
blind zones.  The infrastructure should have the capability to sense whether these blind zones are in fact
occupied by users U2, U3, U4.  
Sensing the \textit{absence} of users in these areas would be communicated  by an I2V  intersection collision avoidance (ICA) message. Upon receiving this message, the ego vehicle would proceed with the assurance that all conflicts are resolved.  If this message is not  received, the vehicle should wait for its signal to change to green.

\section{An intelligent intersection} \label{sec4a}
An intelligent intersection  implements the four-step approach described above.
It has a map of the intersection that includes guideways, movements, conflict zones and blind zones; and sensors
that  detect the occupancy of these blind zones.  The map is constructed offline. The map may be downloaded by connected vehicles and other users.  The intersection software  records the signal phase in order to  calculate SPaT messages in real time,  for example using the algorithms described in \citep{SPATDileep}.   The intersection broadcasts in real time the SPaT message and the occupancy of the blind zones.
These broadcasts are received by connected users of the intersection.  

\begin{figure}[h!]
\centering
\includegraphics[width=6in]{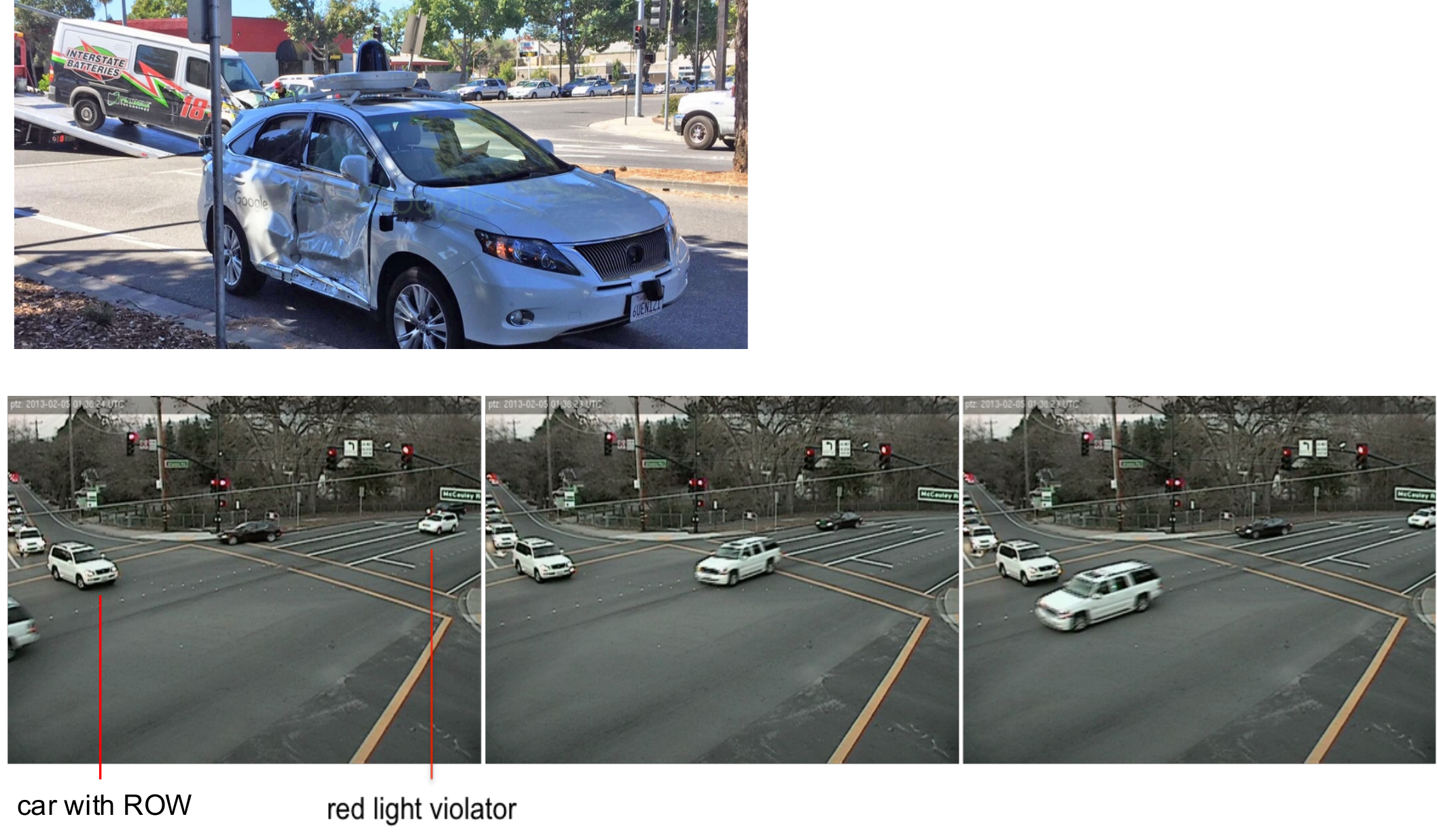}
\caption{Intrusion of red-light violator detected by intersection (source: \citep{HR_TRC}).  Above: picture of a
Google AV crash by a red-light violator.}
\label{fig-fig6b}
\end{figure}

An intelligent intersection can also detect red-light violators.  Figure \ref{fig-fig6b} shows three frames of an intersection video camera.
The camera is triggered by the detection of the intruding vehicle at a high speed traveling into the intersection during a red signal.  The detection took place at least 2s before the vehicle entered the intersection and could warn the vehicle with ROW. Also shown is a picture of a Google AV after a van crashed into it as it entered the intersection 6s into red.  The Google AV could have been programmed to avoid the crash with a 2s warning.

\textbf{Protected intersection.} The intelligent intersection  also functions as a limited but  flexible  protected intersection.  
\begin{figure}[h!]
\centering
\includegraphics[width=6in]{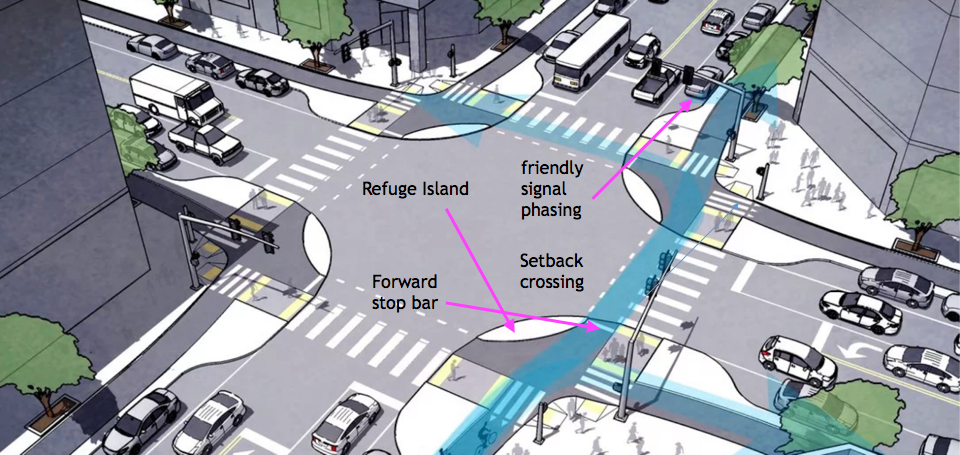}
\caption{Schematic of a protected intersection (https://vimeo.com/86721046).}
\label{fig-fig6a}
\end{figure}
Protected intersections \citep{Falbo} refer to physical modifications designed  to improve the passage of cyclists  through an intersection. Its key  features are (see Figure \ref{fig-fig6a}):
\begin{enumerate}[noitemsep,topsep=0pt]
\setlength{\itemsep}{0in}
\item Insertion of `refuge islands'  to sharpen turning radius of
cars, forcing them to slow down to 5-10 mph when turning right;
\item Special bike lane setback as they cross the intersection;
\item Forward stop bar for cyclists, far ahead of waiting cars;
\item Special cyclist-activated traffic lights;
\item Advance green traffic signals for cyclists, 
\item Turn restrictions for cars, while all turns allowed for cyclists.
\end{enumerate}
Several protected intersections  have been built in the U.S. since 2015.  The first, in Davis, CA cost
\$1M \citep{CityofDavis}.
Not all protected intersections incorporate the special bike signals. In Salt Lake City, UT these signals were not added due to the need to install bicycle detection sensors. Similarly, the protected intersection design in Davis, CA omitted bicycle-friendly signal timing, as it would ``cause backups and could decrease the safety of other parts of the corridor.''

The protected intersection imposes significant mobility cost. As seen in Figure \ref{fig-fig6a} the  right turn pockets have been eliminated so that right turn and through vehicles must share the same lane; the former will block the latter as they wait
to complete the turn, resulting in a significant reduction in the intersection throughput. This reduction is a permanent imposition even when there is no bicycle traffic or when emergency vehicles need to travel quickly.

An intelligent intersection can be enhanced to provide several safety benefits.  Bicyclists or pedestrians could put apps in their smartphones that alert the intersection controller of their location and direction thereby serving as a mobile bicycle or pedestrian sensor.  Knowing how many bicycles there are and their desired turn movements, the controller could adaptively set the duration of the bicycle signal to reduce  backups.  The SPaT calculation could be used to signal to cyclists that they should speed up or slow down to avoid stopping as in the ``Flo'' system introduced in Utrecht \citep{Flo}. 

The cost of an intelligent intersection is relatively small, estimated at \$25K to \$50K, depending on the size of the intersection and the extent of preexisting sensing.  The safety benefits of an intersection upgrade depends on the traffic demand.  From the map of the intersection one can calculate the conflict zones and roughly estimate queues to see how frequently blind zones will occur.  On that basis one can rank intersections and target limited funds.

\section{Preventing the Uber accident with I2V} \label{sec5}

\begin{figure}[h!]
\centering
\includegraphics[width=6in]{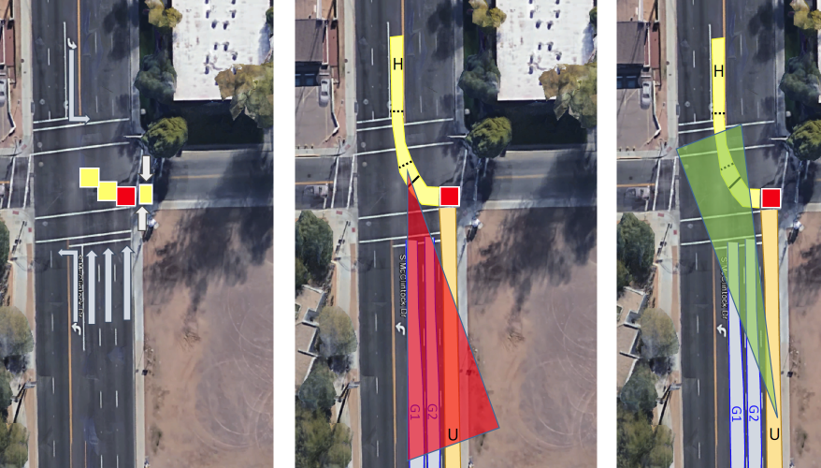}
\caption{Left: the relevant movements and conflict zones; Middle: Guideways G1 and G2 create a blind zone for Honda (H); Right:
G1 and G2 create a blind zone for Uber (U).}
\label{fig-fig7}
\end{figure}
We re-examine the crash summarized in Figure \ref{fig-fig1}, assuming that the intersection is intelligent and that one or both vehicles are connected.  The left panel of Figure \ref{fig-fig7} shows the relevant five vehicle movements and the pedestrian movement on
the crosswalk on the right (phase P6).  The left turn movement from the north (made by the Honda) crosses four conflict zones depicted by squares.  The third conflict zone
(in red) is the one that corresponds to the through movement of the Uber in the rightmost lane. 

The guideways of the three northbound through movements are labeled G1, G2 and U, and the guideway of the left turn
movement is labeled H.  These are shown in the middle panel.  The vehicles in G1 and G2 are stopped so the left-turning
vehicle can see them.  It can also see the pedestrian crossing.  So the only unresolved conflict is with guideway U.  The intersection of H and U is the red conflict zone.  

As shown in the middle panel the stopped vehicles in G1 and G2 cut out the red triangular region from the Honda's field of view.   Hence the Honda would be unable to see the Uber  vehicle if it was more than (say) 50 feet from the red conflict zone.  Thus the region of guidway U upstream of the red conflict zone upto a distance of 150 feet is a blind zone for the
left turning vehicle.  (This calculation is based on the assumption that the Honda can stop in $\tau = 3$s during which time the Uber can travel 150 feet.) 
An advance vehicle detector in the rightmost lane say 200 feet from the stop bar could signal when a vehicle crossed the detector and occupied the blind zone.  The intelligent intersection could broadcast this occupancy message and the Honda would receive it if it were connected, and stop.  But if the Honda was not connected, its driver would not stop and it would be up to the Volvo to prevent the crash.

As shown in the right panel the stopped vehicles in G1 and G2 cut out the green triangular region from the field of view of the automated Volvo.  Hence the region of guideway H from the stop bar to the intersection exit is a blind zone for the Volvo.
The SPaT message from the intelligent intersection would warn the Volvo that its signal will turn yellow several seconds in advance.  Since the Volvo's view of guideway H is obscured, upon receiving the yellow light warning, it would slow down sufficiently as it approached the stop bar so as to see the Honda in time and stop.  Detectors at
the stop bar and at the end of the left turn movement could also determine the presence of the Honda  and the intersection would broadcast a message if no vehicle is detected and the Volvo could continue. However, in the scenario under consideration, the intersection would not send this message, the Volvo would infer that a vehicle (Honda) is occupying the blind zone, slow down and stop.

\section{Illustration of intersection intelligence design} \label{sec5a}
\begin{figure}[h!]
\centering
\includegraphics[width=6.5in]{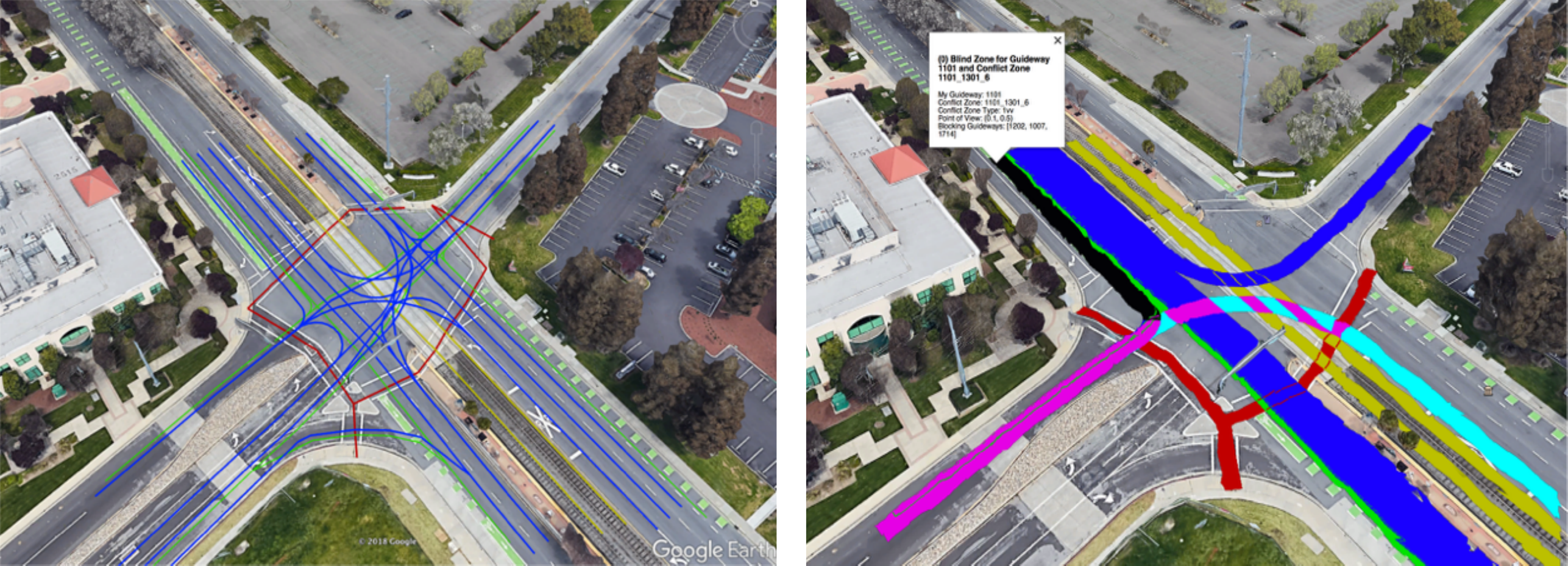}
\caption{Left: Center lines of guideways constructed automatically from the Open Street Map. Right: Actual guideways, conflict zones, and a blind zone of one conflict zone.}
\label{fig-fig9}
\end{figure}
Figure \ref{fig-fig9} illustrates our procedure for automatically constructing guideways, conflict zones, and blind zones of intersections.  We start by  importing a map of the intersection at N First St and Component Drive in San Jose, CA, from the OpenStreetMap database. This map is not shown.  The map gives the lanes of the intersection.  From this we construct the center lines of all guideways. The result is superimposed on a Google Earth map as shown in the image on the left. A guideway's center line starts in the center of an entry lane and ends in the center of an exit lane.  The guideways are obtained by ``thickening'' their center lines as in the image on the right.  Different guideways are shown in different colors. Note the guideways for bicycle lanes, pedestrian crosswalks and light rail.  The conflict zones, which are the intersection of two guideways, and one blind zone in black are shown as in Figure \ref{fig-fig7}.  Details of the code are available in \citep{toolbox}.

\section{Conclusions} \label{sec6}
Vision Zero (VZ) plans and autonomous vehicles (AV) are two significant  approaches to reducing serious injuries and deaths from road accidents.
VZ plans seek to lower the accident rate  through physical redesign of intersections to protect cyclists by separating them from vehicles making   turns, by pedestrian-friendly signal timing such as Leading Pedestrian Intervals, by reducing  the number of lanes, and  lower speed limits for vehicles.  The redesigns are expensive and appear to make cycling safer and more popular.  Rigorous enforcement of VZ rules have safety benefits.   But VZ plans are inflexible and impose significant mobility costs on vehicle drivers.  

Autonomous vehicles (AVs) have  attracted praise from transportation departments and finance from venture capitalists \citep{AV3.0}.  The claim is that AVs will eventually eliminate the 94 percent of crashes involving human error, with no sacrifice of mobility. 
The claim is unverifiable at present.  AVs tested in California reported 40,000 miles of autonomous driving per accident \citep{CADMV_Acc}. 
In 2017, Waymo reported that its safety drivers disengaged autonomous driving once every 5,500 miles. Waymo reports a disengagement when its evaluation process identifies the event as having ``safety significance'' \citep{CADMV}.  These reported rates can be compared with an estimated 500,000 miles per accident in 2015 (6M crashes for 3 trillion VMT, \citep{NHTSAcrash}).

Intersections present a challenging environment both to urban road users and to AV designers.  Unlike VZ plans that rely on infrastructure redesign and AV control programs that rely solely on its on-board sensors, this paper makes the case that both efforts will be more productive if intelligent intersections can provide information that users and AVs lack.  An intelligent intersection can broadcast two crucial information messages.  A SPaT (signal phase and timing) message gives the full signal phase (not just the partial view of the signal that a user or AV has) and an estimate of the time when the signal phase will change.  Together with what  road users and AV sensors can see, SPaT eliminates most possible conflicts. The second  message informs an AV what  its blind spots are and which of them (if any) is occupied by another user.  Computation of the SPaT message requires real time access to the intersection phase.  Calculation of  blind spot occupancy requires real time sensing of well-defined regions of the intersection.  We have argued that these two messages can resolve 
all conflicts within the intersection.

An intelligent intersection has four capabilities.  First, it has a GIS map that describes all the lanes, stop bar locations, and permissible  movements.  An algorithm can then construct the guideways, conflict zones, possible queue obstructions at stop bars and blind spots as outlined in Section \ref{sec5a}.  Second, it has historical and real time signal phase data that record the time of each phase change.  The phase data and the signal timing plan are used to construct the distribution of the remaining duration of the current phase.  This information is needed to compute the SPaT message.  (The phase data itself can be obtained from the controller or its conflict monitor as in \citep{SPATDileep}.) Third, sensors must be installed at strategic locations along guideways upstream of conflict zones to detect whether a blind zone is occupied.  Fourth, the intersection must be able to transmit the SPaT and blind zone occupancy messages to users and AVs.  The easiest way to communicate would be via a smartphone app. Such an app would be automatically triggered by crossing a geo-fence around the intersection; it would inform the intersections of the smartphone of a pedestrian or cyclist; the intelligent intersection would return information or use the presence of the phone to set signal timing.  We estimate an intelligent intersection upgrade to cost about \$25,000, based on the cost of the intersection described in \cite{HR_TRC}.  If intersection traffic data is available one can estimate how frequently blind zones will be occupied and thereby predict the usefulness of an upgrade.  One can rank order intersections by their utility to direct VZ or I2V investment.  In ongoing work we are attempting to carry out this program for San Francisco.

\section*{Acknowledgements} This research was supported by National Science Foundation EAGER
award 1839843, Berkeley Deep Drive, and California Department of Transportation.

\bibliography{/Users/varaiya/Dropbox/varaiya-Main/varaiya/Bib/traffic}
\end{document}